\documentclass[aps,twocolumn,pra,showpacs]{revtex4}

\usepackage{amsmath,amssymb,amsthm,bbm,latexsym}    
\usepackage[dvips]{graphicx}                
\usepackage{subfigure}                  
\usepackage{multirow}                   

\newcommand{\bra}[1]{\left\langle#1\right|}
\newcommand{\ket}[1]{\left|#1\right\rangle}

\newcommand{\expectation}[1]{\left\langle #1\right\rangle}
\newcommand{\modulus}[1]{\left| #1\right|}

\newcommand{\ndash}{\textendash}
\newcommand{\mdash}{\textemdash}
\newcommand{\ie}{{\it i.e. }}

\newcommand{\me}{\mathrm{e}}
\newcommand{\mi}{\mathrm{i}}

\newcommand{\ucl}{\affiliation{Department of Physics and Astronomy, University College London, Gower Street, London WC1E 6BT, United Kingdom}}

\begin{document}

\title{Entanglement creation and distribution on a graph of exchange-coupled qutrits}
\author{Christopher Hadley}\email{c.hadley@ucl.ac.uk}\ucl
\author{Alessio Serafini}\ucl
\author{Sougato Bose}\ucl
\date{\today}

\begin{abstract}
We propose a protocol that allows {\it both} the creation and the 
distribution of entanglement, resulting in two distant parties
(Alice and Bob) conclusively sharing a bipartite Bell state.  The
system considered is a graph of three-level objects (``qutrits'')
coupled by SU(3) exchange operators. The protocol begins with a
third party (Charlie) encoding two lattice sites in unentangled
states, and allowing unitary evolution under time.  Alice and Bob
perform a projective measurement on their respective qutrits at a
given time, and obtain a maximally entangled Bell state with a
certain probablility. We also consider two further protocols, one based on
simple repetition and the other based on successive measurements
and conditional resetting, and show that the cumulative
probability of creating a Bell state between Alice and Bob tends
to unity.
\end{abstract}
\maketitle


The creation and distribution of entanglement are both of immense
importance to the field of quantum information theory.  To date,
much research has taken place into systems performing these tasks
``separately", but in any potential large-scale realization of a
network of quantum computers it would be ideal to have the
capability of performing both these tasks together. By performing
these tasks separately, we mean situations when two particles are
first entangled by some process and then transmitted using a
different process to distant parties through appropriate quantum
channels. For the latter purpose, the transfer (perfect or
otherwise) of quantum states and entanglement through spin chain
channels has been the focus of much recent work. Imperfect (but
good) state transfer has been studied in homogeneous spin chains
\cite{article2003bose, article2004subrahmanyam}, and more recently
it has been shown that pairs of such chains
\cite{article2004burgarth-bose, article2005burgarth-bose,
article2005burgarth-giovanetti-bose} permit perfect state transfer
for large enough time.  Many other schemes for perfect state transfer in spin graph systems have been proposed, relying on engineered couplings
\cite{article2004christandl-datta-ekert-landahl,article2005nikopoulos-petrosyan-lambropoulos,article2005christandl-datta-dorlas-ekert-kay-landahl,article2005yung-bose, article2005kay-ericsson}, state inversion \cite{article2004albanese-christandl-datta-ekert}, graph state generation \cite{article2004clark-mouraalves-jaksch}, multiqubit encoding \cite{article2004osborne-linden,article2004haselgrove}, and spin ladders \cite{article2005li-shi-chen-song-sun}. Moreover, state transfer has recently been studied for harmonic chains \cite{article2005plenio-semiao}, imperfect artificial spin networks \cite{article2005paternostro-palma-kim-falci}, spin rings with flux \cite{article2004bose-jin-korepin} and for many-particle states \cite{article2005li-song-sun}. Related studies have also been undertaken on the dynamical propagation of entangled states
\cite{article2004amico-osterloh-plastina-fazio-palma}, the ``superballistic'' distribution of entanglement \cite{article2005fitzsimons-twamley} and the realization of quantum memories \cite{article2005song-sun,article2005giampaolo-illuminati-dilisi-desiena}. 
Of course, quantum state transfer (entanglement transfer being an automatic corollary) using several systems other than spin chains have also been studied: quantum dot arrays \cite{article2005nikopoulos-petrosyan-lambropoulos,article2004depasquale-giorgi-pagnelli}, Josephson junction arrays \cite{article2005romito-fazio-bruder}, photons in cavity QED \cite{article1997cirac-zoller-kimble-mabuchi} and flying atoms \cite{article2004biswas-agarwal} being just a few examples.

In view of the above, it would be useful to have schemes which
{\em combine} both the creation and distribution of entanglement
as a single process. Recently, such schemes have been proposed in
the context of harmonic chains
\cite{article2004eisert-plenio-bose-hartley,
article2004plenio-eisert-hartley,
article2005paternostro-kim-park-lee, article2005perales-plenio}. In the context of discrete
variable systems, it is a trivial fact to note that if a single
spin is flipped at a site in a spin-$\frac{1}{2}$ chain with homogeneous exchange
couplings, then at a later time two distant spins will be in a
mixed entangled state which has far less than maximal
entanglement. However, we ideally want to distribute {\em maximal}
   entanglement (for example a {\em Bell state}) between two well
   separated sites so that quantum communication schemes such
   as teleportation, dense coding or remote gates for linking separated quantum processors can be perfectly accomplished.
   One way to achieve this aim is to use entanglement distillation \cite{article2000bennett-divincenzo}, but this would require many entangled pairs to be shared in
   parallel, subsequent gates between them and perfectly entangled pairs obtained only in the asymptotic limit from the initial mixed states. Another
   way to achieve this is to use engineered chains
   \cite{article2004yung-leung-bose}. However, quantum chains with
   specific engineered couplings would be harder to produce than
   those with uniform couplings. So, a naturally arising question would be the following: can we design a protocol to create a Bell state between
   separated parties using uniformly coupled chains of quantum
   systems?

Here, we consider a spin graph-based scheme that performs {\it both} the conclusive creation and the distribution of entanglement, avoiding the difficulties of interfacing systems performing these tasks separately.  We propose a protocol, the final result of which shall be that two distant parties (to whom we shall refer as Alice and Bob) share a maximally entangled Bell pair,
\begin{eqnarray}
\ket{\psi^+_{AB}} = \frac{1}{\sqrt{2}} \left[ \ket{+1}_A\ket{-1}_B + \ket{-1}_A\ket{+1}_B\right], \label{bellAB}
\end{eqnarray}
which may then subsequently be used for any quantum informational
task, such as superdense coding or teleportation
\cite{article2000bennett-divincenzo}.  We consider only
maximally entangled states, since we wish to avoid the
distillation and purification that would be required were
non-maximally entangled states produced.

The proposed system differs from many other systems in that the
Hamiltonian is not that of a Heisenberg model, but is
SU(3)-invariant and permutes three states.  The scheme requires
minimal control, requiring only the encoding of two qutrits in a
pure state by a third party, Charlie (a local unitary), a local
projective measurement by both Alice and Bob, and a global
``resetting'' of the lattice. The scheme shares some of the
advantages of the dual rail based scheme
\cite{article2004burgarth-bose, article2005burgarth-bose,
article2005burgarth-giovanetti-bose}, which, although
interpretable as a single SU(3) chain, is only a scheme for
transmitting quantum states or pregenerated entanglement. The
current scheme differs from this from the point of view of being
able to additionally generate the entanglement in the course of
distribution.

\section{The system}
Our system is a graph of ``qutrits'' (three-level objects, the states of which we shall label $\{-1,0,+1\}$) coupled by SU(3) ``swap operators''.  We consider two graphs\mdash a cross and loop\mdash for comparison.

\begin{small}\begin{figure*}    
 \begin{center}\begin{tabular}{cc}
  \includegraphics[width=12cm,angle=0]{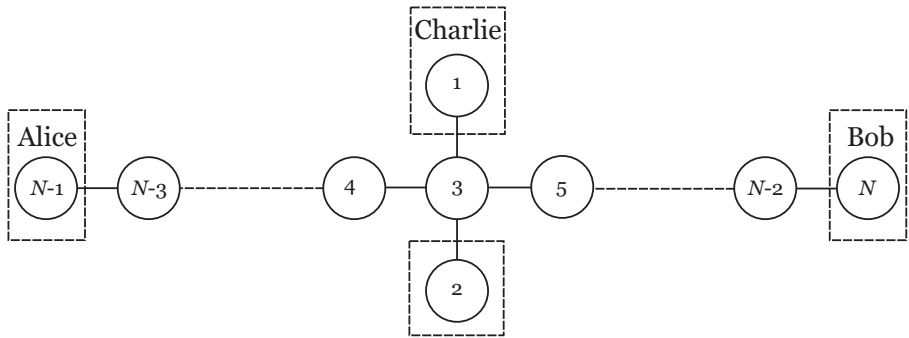}
  \end{tabular}\end{center}
 \caption{The ``cross'' graph of qutrits: each circle represents a three-level quantum system with states $\{-1,0,+1\}$, and the number inside the circle is the vertex index $n\in [1,N]$; lines between qutrits represent the permutation operator (\ref{permutation}).  The qutrits are numbered such that all the even indices are in Alice's arm of the cross, and all the odd indices are in Bob's arm.  The rectangles show the limitations of each person's control.}
 \label{SU3cross}
\end{figure*}\end{small}
\begin{figure}
\includegraphics[width=8cm,angle=0]{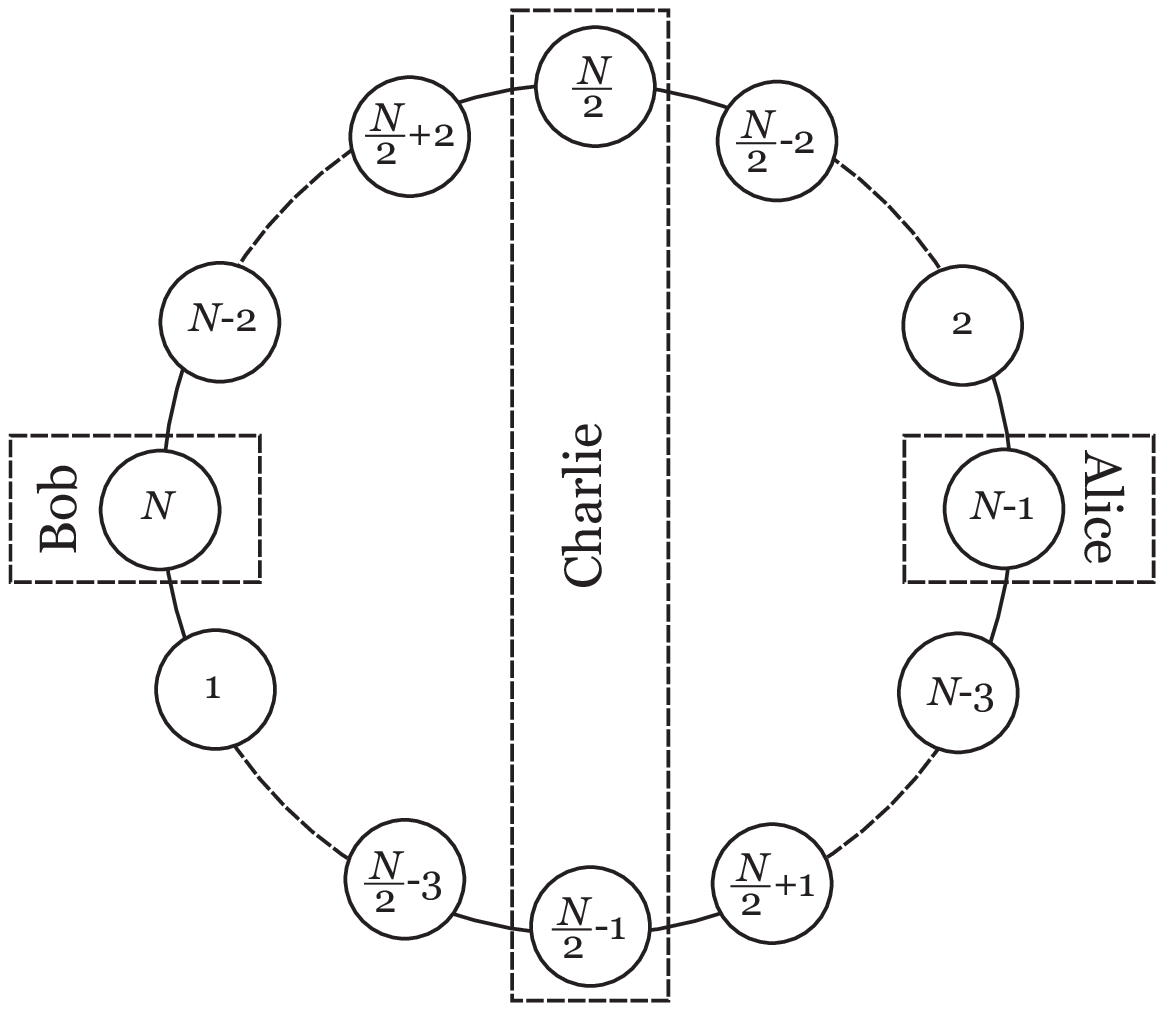}
\caption{The ``loop'' graph of qutrits: as in Figure~\ref{SU3cross} each circle represents a qutrit, labelled $n$.  Note that the Charlie's qutrits could be close together, with Alice and Bob separated by a large distance, or Charlie could be replaced by two distant parties, each in control of one of the qutrits $\frac{N}{2}$ and $\frac{N}{2}-1$.}
\label{SU3loop}
\end{figure}

Let each vertex of the graph be labelled by an index $n\in [1,N]$, and let $\{ S^\alpha_\beta(n) \}$ be the generators of SU(3) at the $n^{\rm th}$ qutrit satisfying the algebra \cite{article2001kiselev-feldmann-oppermann}
\begin{eqnarray}
\left[S^\beta_\alpha (m),S^\rho_\sigma (n) \right] = \delta^m_n \left\{ \delta^\rho_\alpha S^\beta_\sigma (n) - \delta^\beta_\sigma S^\rho_\alpha (n) \label{algebra} \right\}.
\end{eqnarray}
The indices $\alpha, \beta$ refer to the states, and $S^\alpha_\beta(n)$ swaps the states labelled by $\alpha$ and $\beta$ at vertex $n$.  The Hamiltonian is thus \cite{article2001kiselev-feldmann-oppermann, article1988read-sachdev, article1975sutherland}
\begin{eqnarray}
H = J\sum_{\expectation{m,n}}\Pi_{m,n},\label{hamiltonian}
\end{eqnarray}
where the operator
\begin{eqnarray}
\Pi_{m,n} = \sum_{\alpha,\beta}  S^\beta_\alpha (m) S^\alpha_\beta (n)\label{permutation}
\end{eqnarray}
permutes the states at vertices $m$ and $n$ and the sum is taken
over all neigbouring vertices $m, n$ \footnote{The sign of $J$ is
unimportant, as this only sets the energetic ordering of the
states.} and all states $\alpha,\beta$.

A greatly important point to notice is that the generators may be given either bosonic \cite{article1988arovas-auerbach} or fermionic \cite{article2001kiselev-feldmann-oppermann} representations in terms of creation and annihilation operators according to $S^\alpha_\beta (n) = c^{\dag\alpha}(n) c_\beta(n)$.  The physics is representation independent, giving rise to a variety of potential physical implementations.  Indeed, possible realizations of SU(3)-invariant Hamiltonians include optical lattices \cite{article2005osterloh-baig-santos-zoller-lewenstein}, trapped ions \cite{article2005klimov-guzman-retamal-saavedra, article2005mchugh-twamley} and quantum dots \cite{article2005onufriev-marston}.

The state of the whole system may be described in terms of basis vectors $\ket{\psi} = \ket{\psi_1}\otimes\cdots\otimes\ket{\psi_N}$, residing in a Hilbert space of dimension $3^N$, where $\ket{\psi_n}$ is the state at the $n^\mathrm{th}$ site.  However, since the Hamiltonian merely permutes states, the numbers of $+1$ and $-1$ excitations are individually conserved; thus we may describe the state in terms of a smaller basis.  We shall here consider the case where there is always one qutrit in either of the states $\pm 1$, and thus for convenience use the compact basis $\{\ket{i,j}\}_{i\ne j=1}^N$, where $i,j$ are respecively the indices of the lattice sites where the states $+1$ and $-1$ reside; this basis has size $^N P_2 = N(N-1)$.

\section{Protocol: one measurement}
\label{protocolsimple}
Initially, each lattice site is set to the state $\ket{0}$.  For
the cross (Figure \ref{SU3cross}), Charlie has control of sites
$1$ and $2$, and encodes these in the state $\ket{+1}_1\ket{-1}_2$
(or equivalently $\ket{1,2}$ in the reduced basis).  For the
loop (Figure \ref{SU3loop}), Charlie has control of vertices
$\frac{N}{2}$ and $\frac{N}{2}-1$ and encodes these in the similar
state $\ket{\frac{N}{2},\frac{N}{2}-1}$.

The system is then allowed to evolve under the Hamiltonian
(\ref{hamiltonian}). Under this evolution, as it only swaps states
of neighbouring qutrits, the system is always in a state with
exactly one qutrit in $\ket{+1}$, one in $\ket{-1}$ and
the remainder in $\ket{0}$. After a given time Alice and Bob
perform at their respective qutrits the composite, local
projective measurement
\begin{eqnarray}
M &=& \left[ \ket{+1}_A\bra{+1}_A+\ket{-1}_A\bra{-1}_A \right]\nonumber\\
&\otimes&\left[ \ket{+1}_B\bra{+1}_B+\ket{-1}_B\bra{-1}_B \right], \label{measure}
\end{eqnarray}
which effectively tests for the Bell state $\ket{\psi^+_{AB}}$,
since the terms $\ket{\pm 1}_A\bra{\pm 1}_A\otimes\ket{\pm
1}_B\bra{\pm 1}_B$ always give null results.  This allows the
presence of a {\it global} state to be tested through {\it local}
measurements. Each of the parentheses of $M$ performs a {\em coarse
grained measurement} at either Alice or Bob's qutrit which differentiates between states $\ket{\pm
1}$ and $\ket{0}$, but does not distinguish between $\ket{+1}$ and
$\ket{-1}$; \ie it gives the {\em same} eigenvalue $+1$
for {\em both} outcomes $\ket{+1}$ and $\ket{-1}$, while it gives a
different eigenvalue $0$ for the outcome $\ket{0}$. We do not go
into percise details such a coarse-grained measurement but
only mention the fact that it is allowed by quantum mechanics. The
precise mechanism may vary from one physical implementation of our
protocol to another. In optical lattices, for example, if three
internal atomic levels are being used as $\ket{0}$, $\ket{+1}$ and
$\ket{-1}$, then by applying a laser of appropriate frequency and
polarization Alice or Bob can {\em selectively} send an atom in the
state $\ket{0}$ to an unstable excited state. When this state
spontaneously decays (rather immediately), the fluorescence will
tell us that the atom was in the state $\ket{0}$. The absence of fluorescence
will imply that the atom was in either of the states $\ket{+1}$ or
$\ket{-1}$ but not reveal whether it was actually $\ket{+1}$ or
$\ket{-1}$. After the measurement, Alice and Bob perform classical
communication to compare measurement outcomes. If both have
positive measurements (\ie both of their measurements register
$\ket{\pm 1}$, Alice and Bob conclusively \footnote{By
``conclusively'', we mean when success occurs, Alice and Bob are
aware of this.} share the state (\ref{bellAB}); if the
wave function in the $\{\ket{i,j}\}$ basis is $\sum_{i \ne j =
1}^{N} a_{i,j} \ket{i,j}$ immediately before the measurement this
occurs with a probability
\begin{eqnarray}
p^{(1)}(t) = \frac{1}{2} \modulus{ a_{N,N-1} + a_{N-1,N} }^2.\label{onemeasureprobsuccess}
\end{eqnarray}

We have calculated numerically this probability for both the cross and the loop, for various sizes $N$ of these.  The probability is plotted against time in Figure \ref{resultsonemeasure}.

Results for the cross show a characteristic initial peak shortly after the state for all $N\ne 5$.  This would be the optimum time to measure.  The special case $N=5$ arises because of the symmetry of the system.  We later discuss methods of improving the success probability.

For the loop, there is less of a characteristic pattern, although there remains a large peak.  The simplest case $N=4$ has a periodic peak probability of 50\%.

For both graphs, the peak probability decreases with $N$ (see Figure \ref{resultsonemeasureN}).

\begin{small}\begin{figure*}    
 \begin{center}\begin{tabular}{ccc}
  \subfigure[Cross: $N=5$]{\includegraphics[width=4cm,angle=-90]{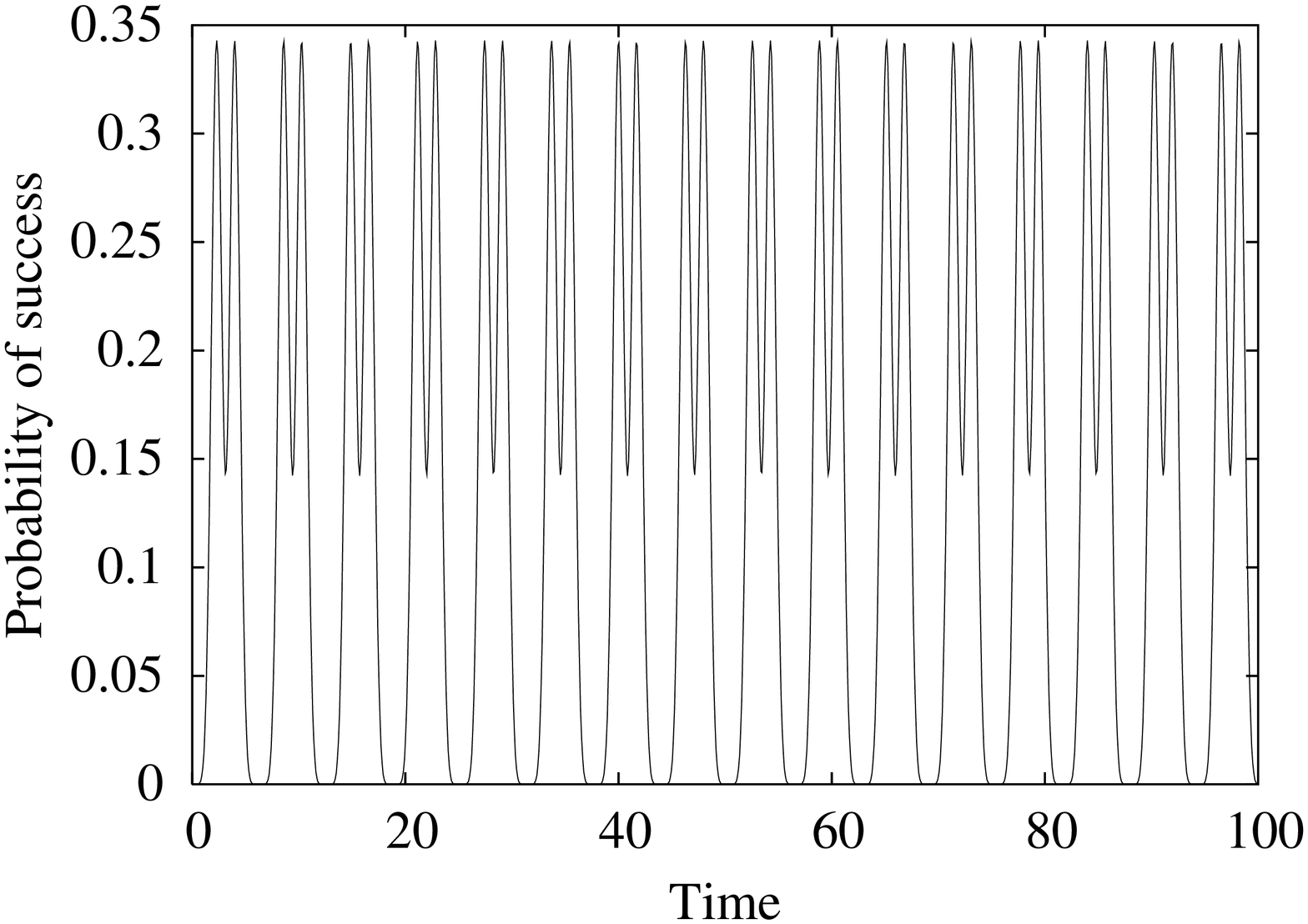}\label{cross5}}
   &
   \subfigure[Cross: $N=13$]{\includegraphics[width=4cm,angle=-90]{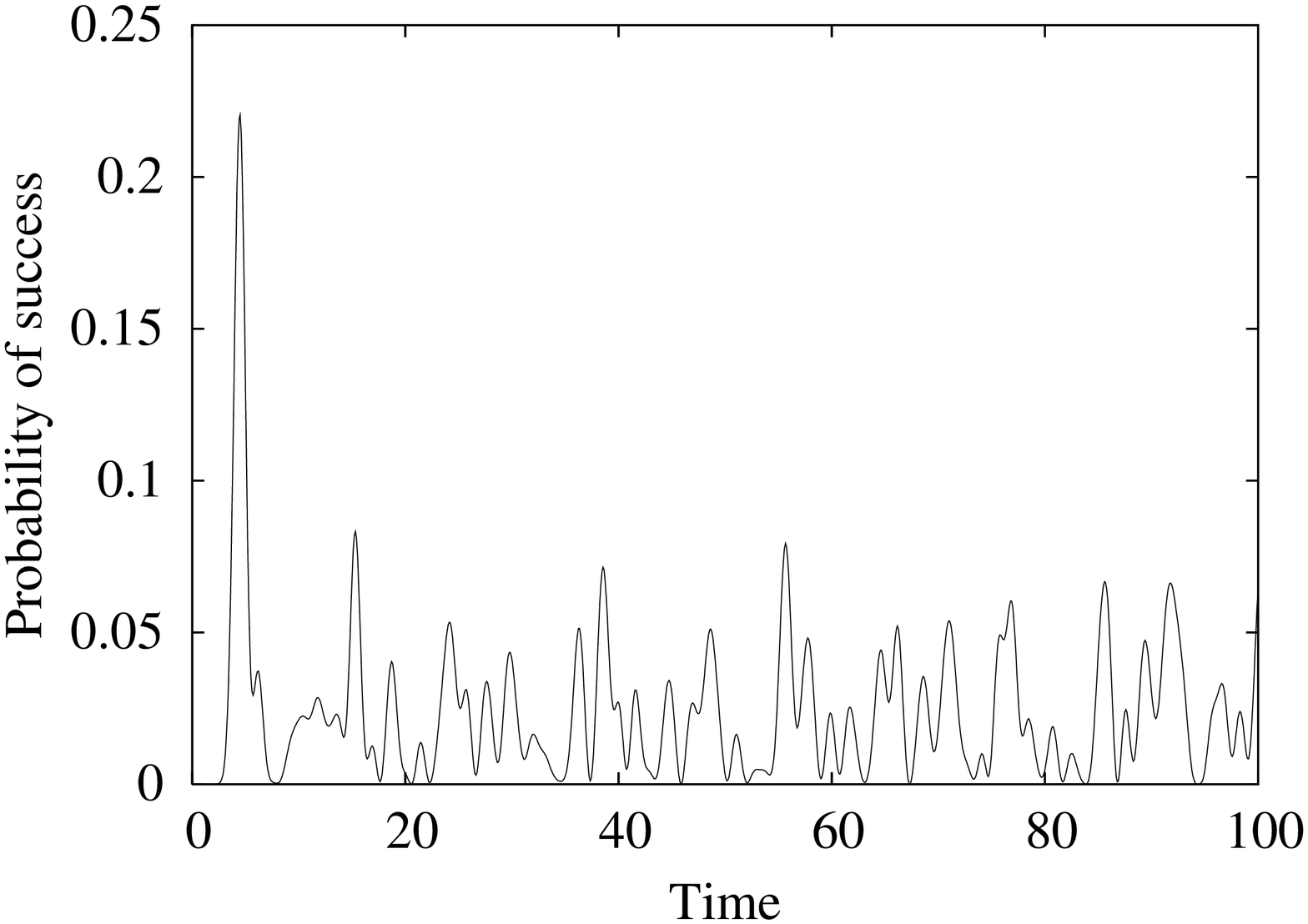}\label{cross13}}
   &
   \subfigure[Cross: $N=19$]{\includegraphics[width=4cm,angle=-90]{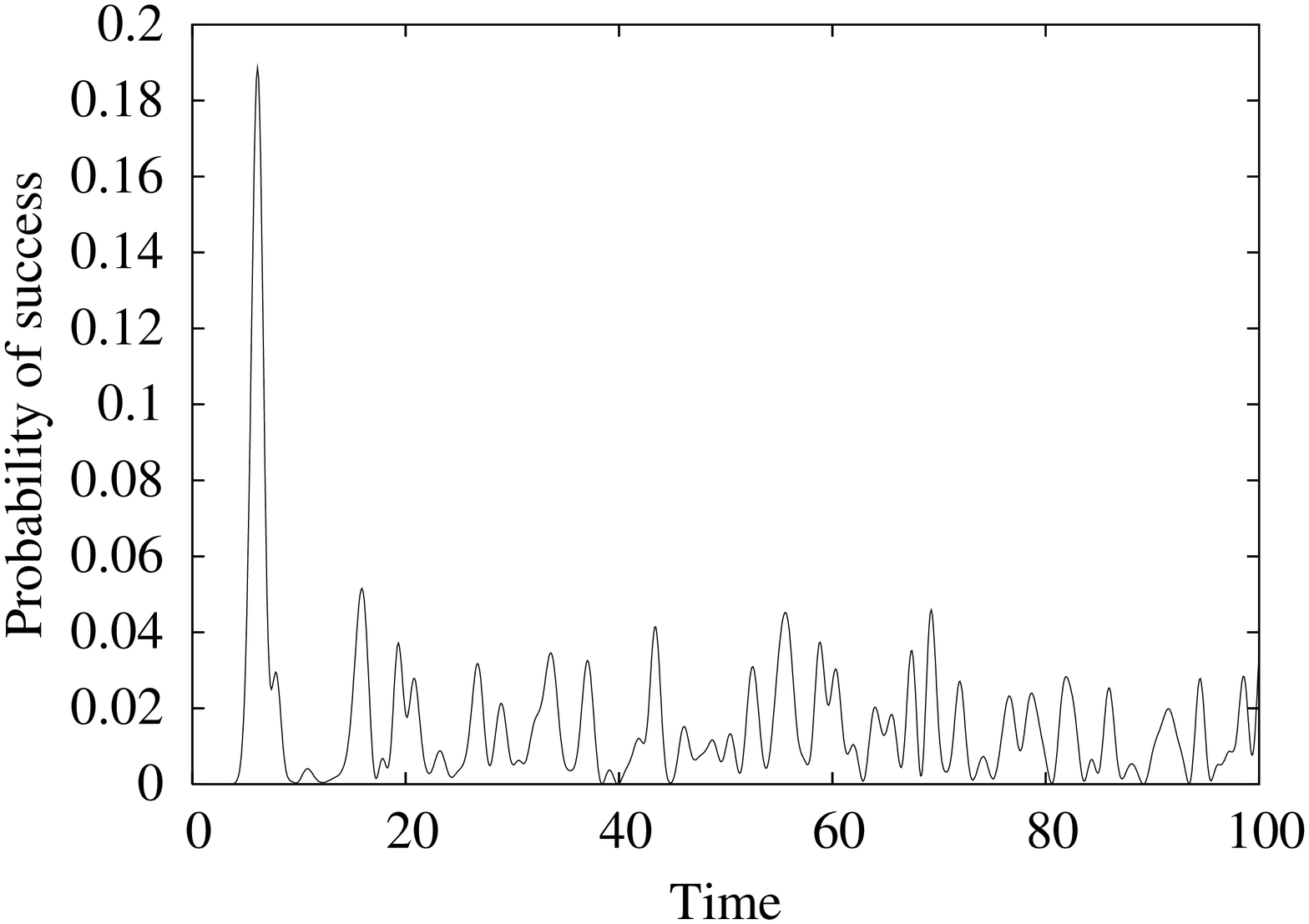}\label{cross19}}
   \\
   \subfigure[Loop: $N=4$]{\includegraphics[width=4cm,angle=-90]{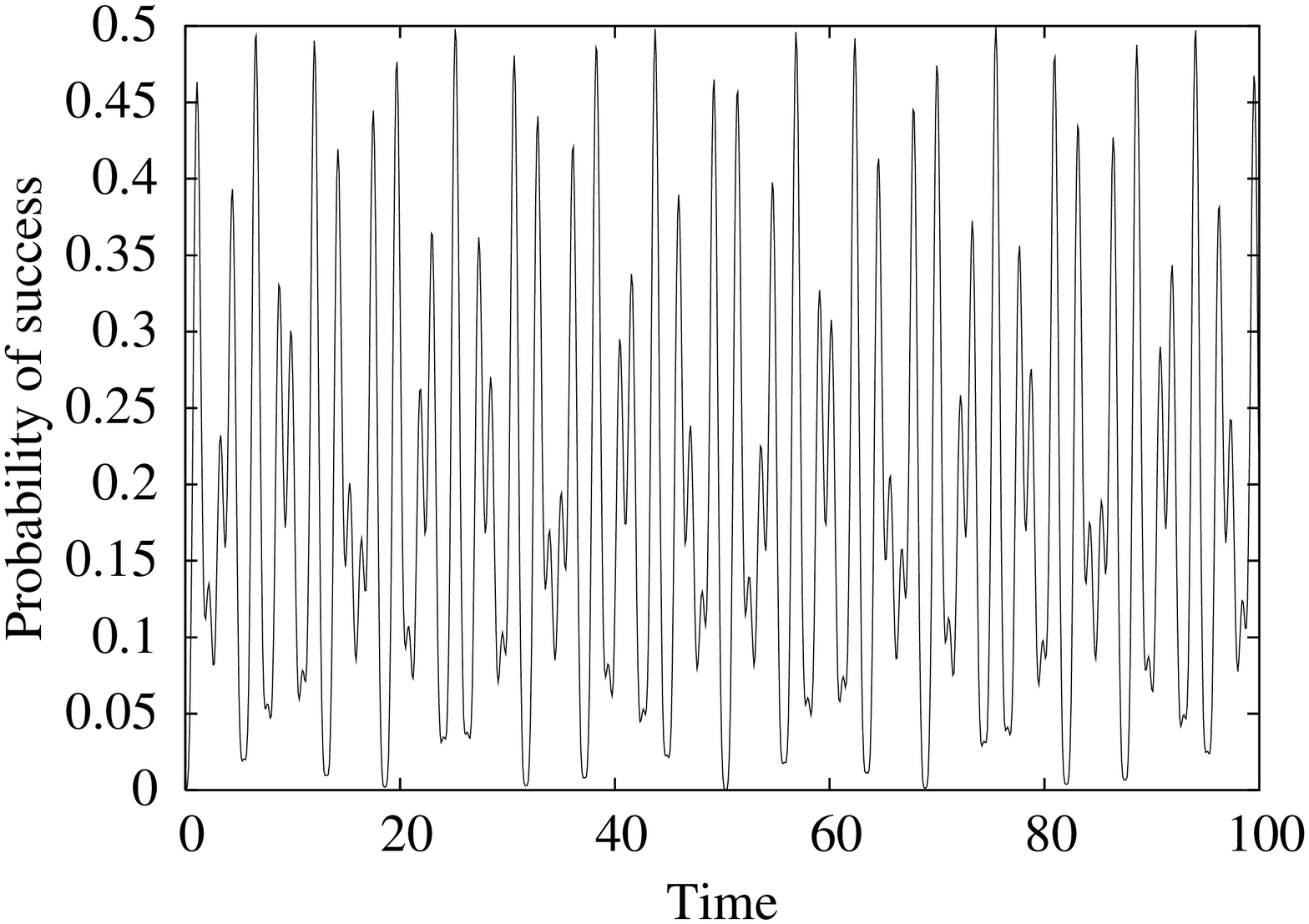}\label{loop4}}
   &
   \subfigure[Loop: $N=8$]{\includegraphics[width=4cm,angle=-90]{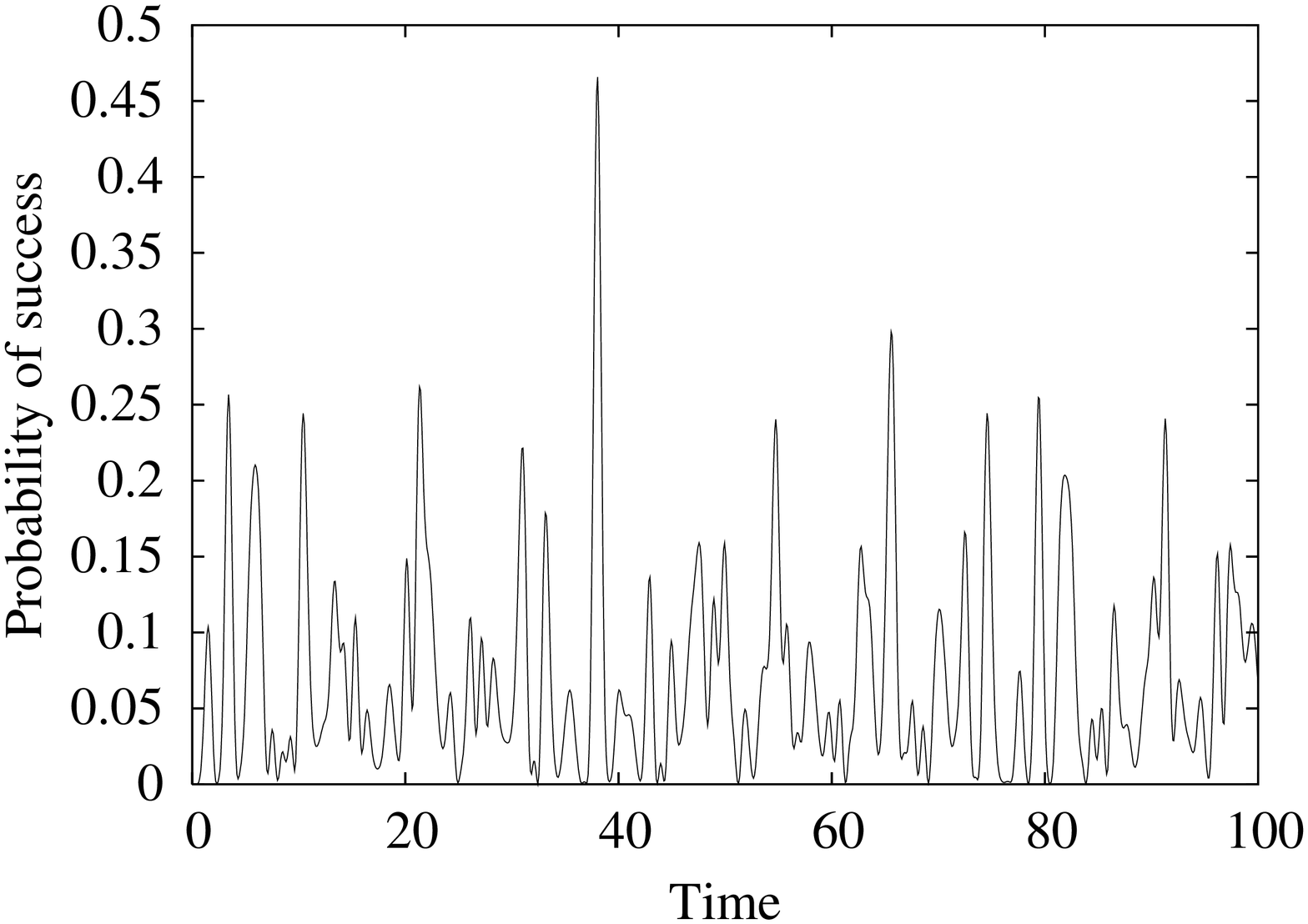}\label{loop8}}
   &
   \subfigure[Loop: $N=16$]{\includegraphics[width=4cm,angle=-90]{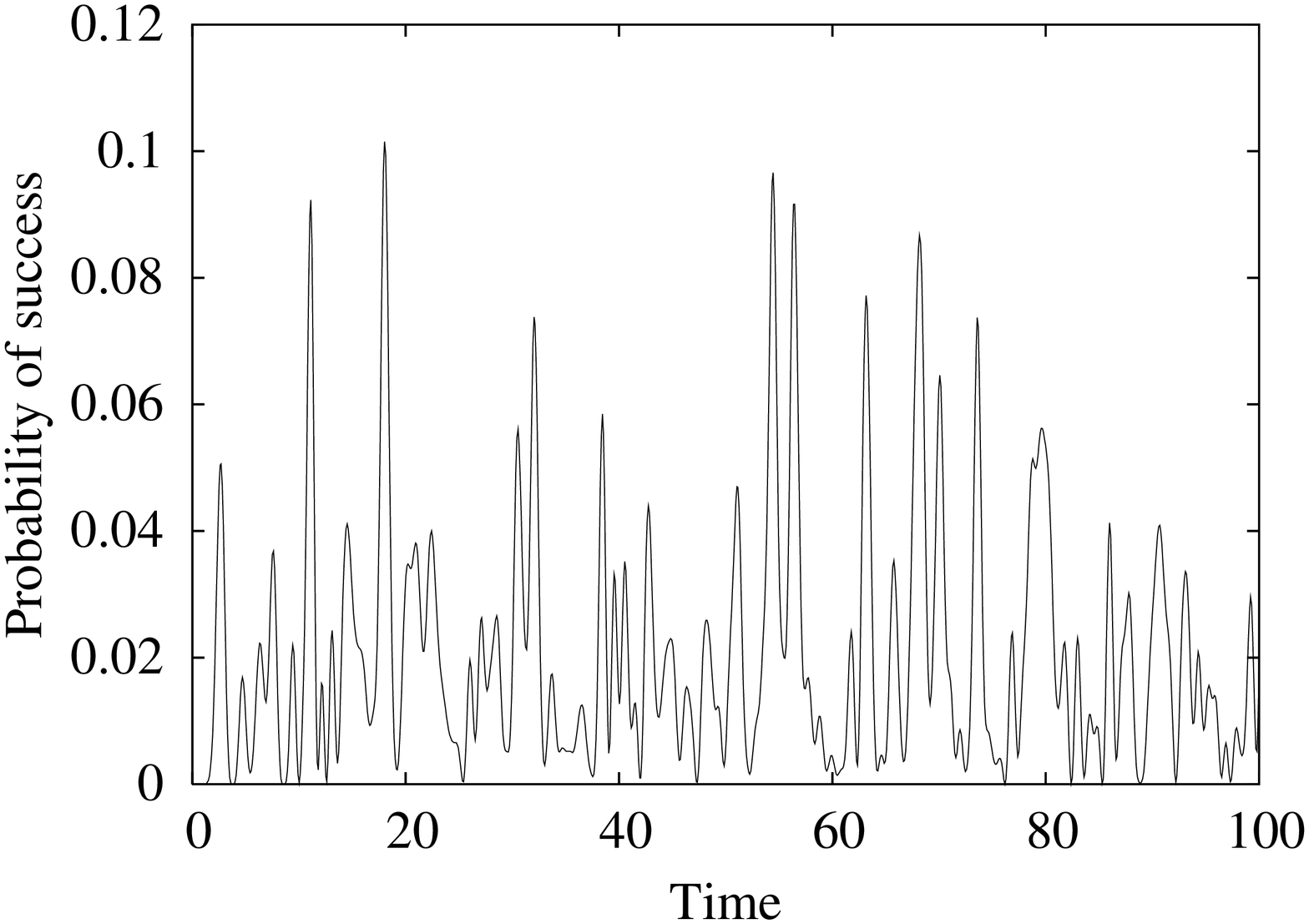}\label{loop16}}
  \end{tabular}\end{center}
 \caption{Probability of success [\ie measurement (\ref{measure}) resulting in Alice and Bob sharing a Bell state (\ref{bellAB})] as a function of time $t$ for various $N$. The time is in units of $\hbar J^{-1}$.}
 \label{resultsonemeasure}
\end{figure*}\end{small}

\begin{small}\begin{figure}
 \begin{center}\begin{tabular}{c}
   \subfigure[Cross]{\includegraphics[width=5cm,angle=270]{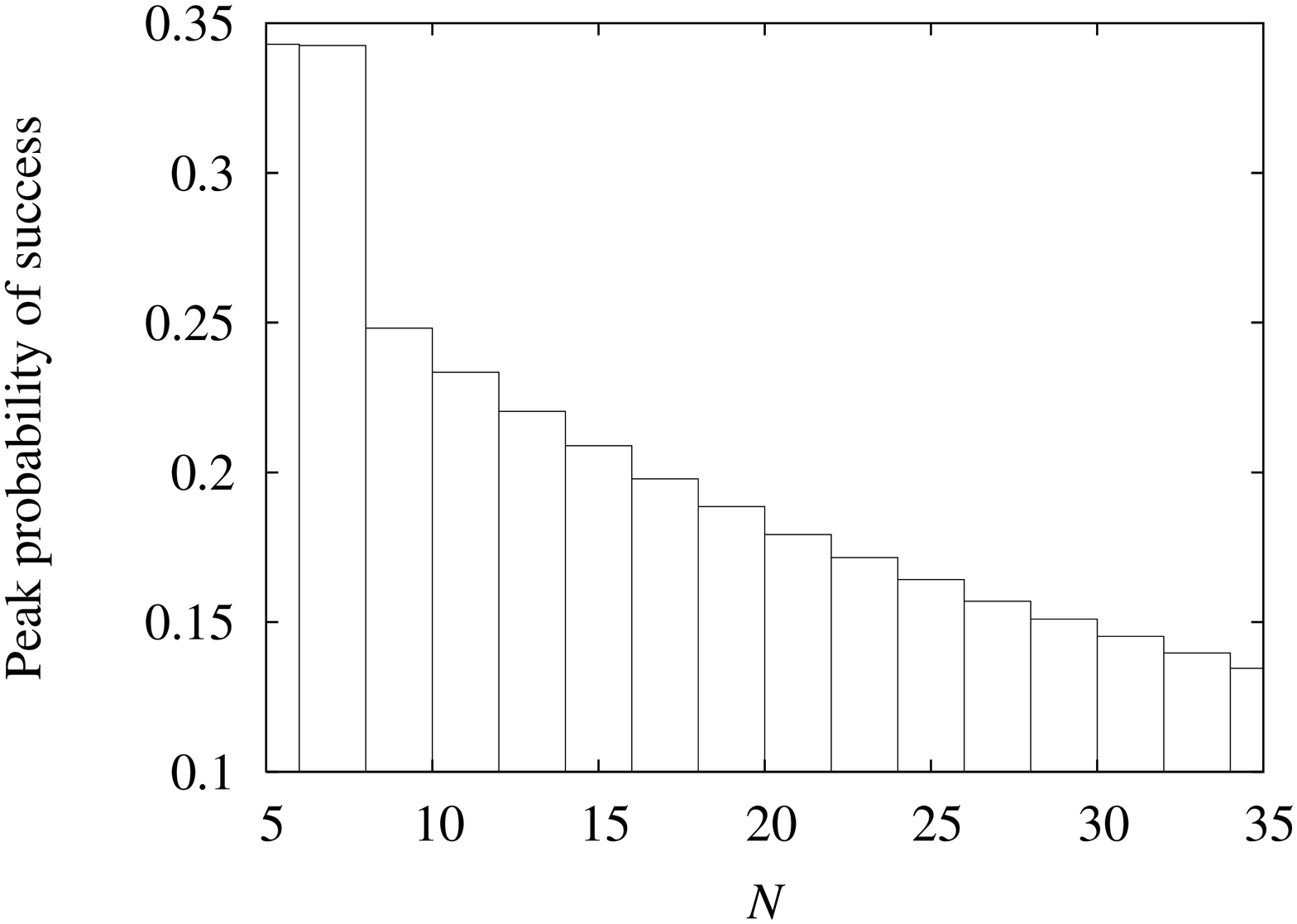}\label{fmaxcross}}
   \\
   \subfigure[Loop]{\includegraphics[width=5cm,angle=270]{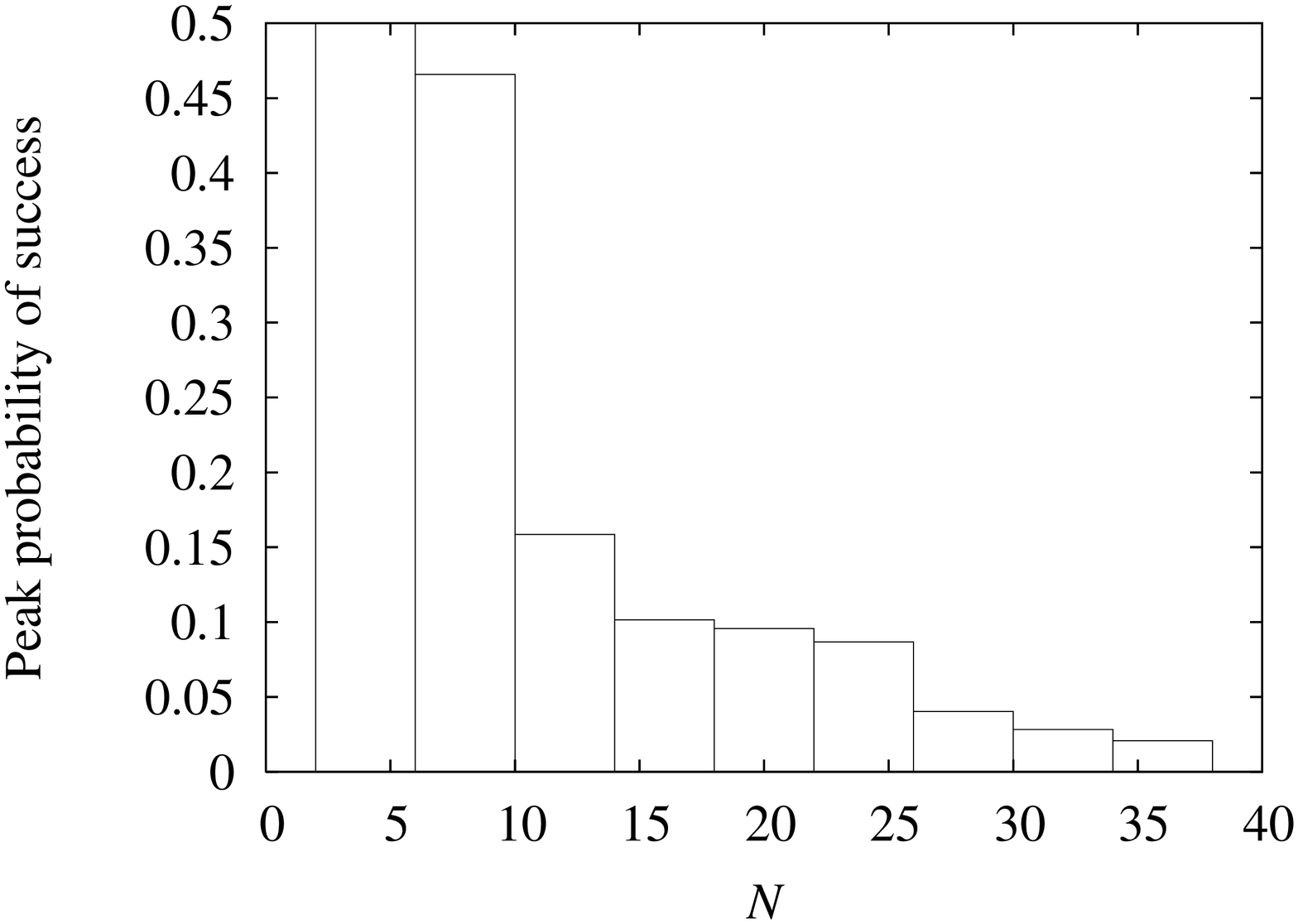}\label{fmaxloop}}

  \end{tabular}\end{center}
 \caption{Peak probability of success as a function of $N$ for one measurement}
 \label{resultsonemeasureN}
\end{figure}\end{small}

\section{Protocol: repeated measurements I}
\label{protocolrepeatedI}
In order to improve the probability of success, one can repeat the protocol many times until success occurs.  If the peak probability of success is $p$, the cumulative probability after $n$ repetitions of this protocol is
\begin{align}
P(n) = \sum_{k=1}^n p(1-p)^{k-1}.
\end{align}
Since this is a geometric progression, it is clear that $\lim_{n\to\infty}P(n)=1$.  The question is how quickly this converges.  In Tables \ref{resultstablecross} and \ref{resultstableloop} we have calculated the number of measurements required to obtain success with a probability of at least $0.90$, $0.95$ and $0.99$ for the cross and loop respectively, when we take measurements at the peak of the probability.  In general it is clear that to obtain success with probability in excess of some $q$, the number of measurements $n$ must satisfy
\begin{align}
n\ge \frac{\log(1-q)}{\log(1-p)}.
\end{align}

\begin{small}
\begin{table}

\begin{tabular*}{0.75\columnwidth}{@{\extracolsep{\fill}}c||c|c|c}
   $N$ & \multicolumn{3}{c}{$q$} \\
  \hline
   & $0.90$ & $0.95$ & $0.99$ \\
  \hline\hline
    5   &    6   &    8   &   11   \\
    7   &    6   &    8   &   11   \\
    9   &    9   &   11   &   17   \\
   11   &    9   &   12   &   18   \\
   13   &   10   &   13   &   19   \\
   15   &   10   &   13   &   20   \\
   17   &   11   &   14   &   21   \\
   19   &   12   &   15   &   23   \\
   21   &   12   &   16   &   24   \\
   23   &   13   &   16   &   25   \\
   25   &   13   &   17   &   26   \\
   27   &   14   &   18   &   27   \\
   29   &   15   &   19   &   29   \\
   31   &   15   &   20   &   30   \\
   33   &   16   &   20   &   31   \\
   35   &   16   &   21   &   32   \\
\end{tabular*}
\caption{Cross: number of measurements required to obtain at least desired probability of success $q$ for various $N$.}\label{resultstablecross}

\begin{tabular*}{0.75\columnwidth}{@{\extracolsep{\fill}}c||c|c|c}
  $N$ & \multicolumn{3}{c}{$q$} \\
  \hline
   & $0.90$ & $0.95$ & $0.99$ \\
  \hline\hline
    4   &    4   &    5   &    7   \\
    8   &    4   &    5   &    8   \\
   12   &   14   &   18   &   27   \\
   16   &   22   &   28   &   44   \\
   20   &   23   &   30   &   46   \\
   24   &   26   &   33   &   51   \\
   28   &   57   &   73   &   113   \\
   32   &   81   &   105   &   161   \\
   36   &   110   &   143   &   220   \\
\end{tabular*}
\caption{Loop: number of measurements required to obtain at least desired probability of success $q$ for various $N$.}\label{resultstableloop}

\end{table}
\end{small}

We see that the probability of success converges relatively
quickly; however, the system needs to be reset at each stage
after an unsuccessful measurement.  If the three levels of each
qutrit are represented by hyperfine levels of atoms, with energy
of atoms in $\ket{0}$ lower in a magnetic field than atoms in
$\ket{\pm 1}$, then the resetting can be achieved by applying a
uniform magnetic field to the system and bringing it to its ground
state (this resetting process, of course, is not unitary). Optical
pumping, as used to initialize quantum registers in optical
lattices, can also be used \cite{article2004schrader-dotsenko-etal}. In other
physical implementations, cooling the system to a ground state
could be possible.

Note that in principle, resetting could be achieved through local
actions of Alice and Bob.  Upon unsuccessful measurement, Alice
and Bob continue to measure periodically, and when one receives an
excitation, he or she swaps the qutrit out of the system for a new
qutrit in the state $\ket{0}$.  Alice and Bob continue until two
excitations have been removed in this way; they then know that the
whole lattice is in the initial state.

\section{Protocol: repeated measurements II}
\label{protocolrepeatedII} In order to reduce the number of times
the system is reset, we now propose another protocol.  First, let
us consider in more detail what happens when the measurement
(\ref{measure}) is applied.  This measurement distinguishes
between $\ket{\pm 1}$ and $\ket{0}$ at each site.  There are thus
four possible outcomes: (i) both measurements are negative, giving
the state $\ket{0}_A\ket{0}_B$, with the excitations remaining
elsewhere in the system; (ii) Alice's measurement is positive, and
Bob's negative, giving the state $\ket{\pm 1}_A\ket{0}_B$
\footnote{The measurement cannot distinguish these states.}; (iii)
Alice's is negative, and Bob's positive, giving the state
$\ket{0}_A\ket{\pm 1}_B$; (iv) both are positive, and Alice and
Bob share the state (\ref{bellAB}).

In each case, the resultant overall wave function will be different.  If the wave function in the $\{\ket{i,j}\}$ basis is $\sum_{i \ne j = 1}^{N} a_{i,j} \ket{i,j}$ immediately before the measurement, these will be for the both cross and the loop, respectively:
\begin{align}
\ket{\psi_1} &= \frac{\sum_{i \ne j = 1}^{N-2} a_{i,j} \ket{i,j}}{\sqrt{\sum_{i \ne j = 1}^{N-2} |a_{i,j}|^2}};\\
\ket{\psi_2} &= \frac{ \sum_{j = 1}^{N-2} (a_{N-1,j} \ket{N-1,j} + a_{j,N-1} \ket{j,N-1} ) }{ \sqrt{ \sum_{j = 1}^{N-2} (|a_{N-1,j}|^2 + |a_{j,N-1}|^2 )} };\\
\ket{\psi_3} &= \frac{ \sum_{j = 1}^{N-2} (a_{N,j} \ket{N,j} + a_{j,N} \ket{j,N} ) }{ \sqrt{ \sum_{j = 1}^{N-2} (|a_{N,j}|^2  + |a_{j,N}|^2 )} };\\
\ket{\psi_S} &= \frac{1}{\sqrt{2}} \left[ \ket{N-1,N} + \ket{N,N-1} \right];
\end{align}
where Alice's qutrit is at vertex $N-1$ and Bob's at $N$, and of course $\sum_{i\ne j =1}^N \modulus{a_{i,j}}^2 = 1$.

If the measurement is unsuccessful, we end up with one of the states $\ket{\psi_{1\ndash3}}$.  Since the measurement has not totally destroyed the amplitude of the excitations existing in the system, we may take another measurement some time later.

However, it is difficult numerically to consider simultaneously the separate evolutions of the states $\ket{\psi_{1\ndash 3}}$.  Since the states $\ket{\psi_{2,3}}$ are asymmetric (\ie Alice and Bob do not have the same local states), let us consider taking repeated measurements only on the outcome $\ket{\psi_1}$, which possesses the same symmetry at the target state and thus seems most likely that it will lead to this.

Our protocol now is that at each measurement, if the outcomes $\ket{\psi_{2,3}}$ occur, Alice, Bob and Charlie reset all qutrits to $\ket{0}$ and start again.  All possible outcomes are represented diagrammatically in Figure \ref{outcomes}.

\begin{small}\begin{figure*}	
 \begin{center}
  \includegraphics[width=12cm,angle=0]{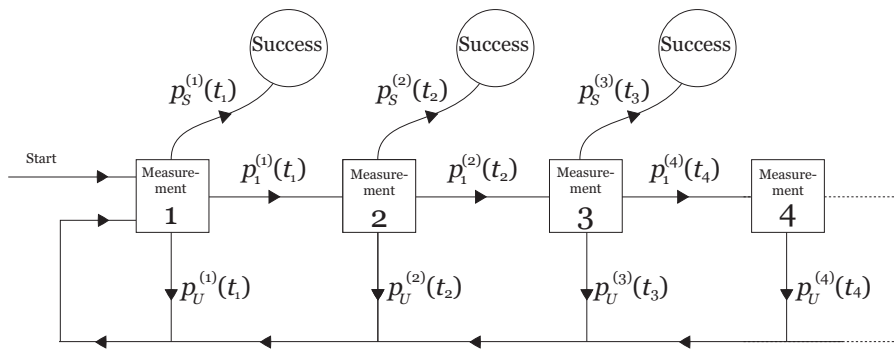}
  \end{center}
 \caption{Diagrammatic representation of protocol discussed in Section \ref{protocolrepeatedII}.  Alice and Bob's route through the diagram starts on the left.  At each measurement, there are three types of outcome: success, a state $\ket{\psi_1}$ (upon which they carry on, represented by rightwards arrows), or state $\ket{\psi_{2,3}}$ (upon which they reset and start again, represented by downwards arrows).}
 \label{outcomes}
\end{figure*}\end{small}

It is clear that the cumulative probability of success occurring by the $n^{\rm th}$ measurement without having to restart is then
\begin{eqnarray}
\bar{P}_n(t_1,\ldots ,t_n) = p^{(1)}_S(t_1) + \sum_{j=2}^n p^{(j)}_S(t_j) \prod_{i=1}^{j-1} p^{(i)}_1(t_i)\label{cumulative}
\end{eqnarray}
for $n$ measurements at times $t_{1\ndash n}$.  Alice and Bob's strategy should be to attempt to maximise this with respect to $t_{1\ndash n}$, where $p^{(i)}_k (t_i)$ is the probability of outcome $k \in \{1,2,3,S\}$ at the $i^{\rm th}$ measurement.

When considering only one measurement, the optimum strategy was to measure when the probability of success was at a peak.  Here however, three strategies naturally present themselves when considering at what time each successive measurement should be taken: (i) measure when the probability of success is at a peak, as above; (ii) measure when the probability of states $\ket{\psi_{2,3}}$ is at a minimum, so we are minimizing the amount of wavefunction we are ``throwing away''; (iii) measure when the difference between the probability of success and the probability of receiving the states $\ket{\psi_{2,3}}$ is maximized.

There is an important point to notice here: since we are taking measurements when the one-shot probability of success is at a maximum, the time of the $n^{\rm th}$ measurement depends on the route taken through all the possibilities in Figure \ref{outcomes}.  Any problems caused by this could be rectified by taking measurements at regular time intervals, such that $t_k = k\tau$ for all $k$.  However, we shall continue to optimise the probabilities at each stage according to the three strategies (i)\ndash(iii), since taking measurements at regular intervals may cause some measurements to be taken at troughs in the success probability, thus causing the system to require more measurements.

A moment's thought should convince one that the cumulative probability of success will tend towards unity with increased number of measurements, since we should receive the state eventually.  Let us denote single event probabilities by lower-case $p$'s, and joint probabilities with upper-case $P$'s.  Now, we reset the state on measurement of either of the states $\ket{\psi_{2,3}}$.  The probability of receiving either of these and thus requiring restarting at the $i^{\rm th}$ measurement is $p_U^{(i)}(t_i) = p_2^{(i)}(t_i) + p_3^{(i)}(t_i)$.

The total probability of receiving the state by the $n^{\rm th}$ measurement is then
\begin{widetext}
\begin{align}
P_n(t_1,\ldots,t_n) = \bar{P}_n(t_1,\ldots,t_n) + p_U^{(1)}(t_1)P_{n-1}(t_2,\ldots,t_n) + &\sum_{j=2}^{n-1} \left\{ \prod_{i=1}^{j-1}p_1^{(i)}(t_i) \right\} p_U^{(j)}(t_j) P_{n-j}(t_{j+1},\ldots,t_n)\\
= \bar{P}_n(t_1,\ldots,t_n) + p_U^{(1)}(t_1)P_{n-1}(t_1,\ldots,t_{n-1}) + &\sum_{j=2}^{n-1} \left\{ \prod_{i=1}^{j-1}p_1^{(i)}(t_i) \right\} p_U^{(j)}(t_j) P_{n-j}(t_1,\ldots,t_{n-j}),
\end{align}
\end{widetext}
where we have used the fact that $P_{n-j}(t_{j+1},\ldots,t_n) = P_{n-j}(t_1,\ldots,t_{n-j})$.  This formula for the total cumulative probability can be built up iteratively, since one can find $P_{n+1}(t_1,\ldots,t_{n+1})$ from knowing $P_1(t_1),\ldots,P_n(t_1,\ldots,t_n)$.  Note that the $\bar{P}_n$ are already known from previous numerical calculations in Section \ref{protocolsimple}.

For this protocol to work arbitarily well, we would like it to be the case that
\begin{eqnarray}
\lim_{n\to\infty} P_{n}(t_1,\ldots,t_n) = 1.\label{prooflim}
\end{eqnarray}
We now give a simple argument that this is indeed the case.  Let $m$ be the number of times the system has to be reset, and $n$ the total number of measurements taken, and make the assumption that if the probability $p_U$ of having to reset is zero at some time there exists at least one subsequent time at which is non-zero, such that as $n\to\infty$, so too does $m\to\infty$.  It is then possible to say that the probability of success occurring between the $j^{\rm th}$ and $(j+1)^{\rm th}$ resettings is always greater than or equal to the probability $p$ of the initial peak, since measuring again can only increase or have no effect on the cumulative probability.  This then implies
\begin{align}
\lim_{n\to\infty}P_n(t_1,\ldots,t_n) \ge \lim_{m\to\infty} \sum_{k=1}^m p(p-1)^{k-1} = 1,\label{proofineq}
\end{align}
and thus (\ref{prooflim}) is satisfied.
\begin{small}
\begin{table*}
\begin{tabular*}{10cm}{@{\extracolsep{\fill}}c|c c c|c c c}
 & \multicolumn{3}{c|}{Protocol II} & \multicolumn{3}{c}{Protocol I} \\
 Measurement & $N=5$ & $N=7$ & $N=9$ & $N=5$ & $N=7$ & $N=9$ \\
 \hline
 1 & 0.3429 & 0.3426 & 0.2482   &   0.3429 & 0.3426 & 0.2482 \\
 2 & 0.5294 & 0.5091 & 0.3966   &   0.5682 & 0.5679 & 0.4735 \\
 3 & 0.6667 & 0.5937 & 0.4794   &   0.7162 & 0.7160 & 0.6216 \\
 4 & 0.7620 & 0.6614 & 0.5344   &   0.8136 & 0.8133 & 0.7189 \\
 5 & 0.8280 & 0.7061 & 0.5737   &   0.8775 & 0.8772 & 0.7828 \\
 6 & 0.8741 & 0.7461 & 0.6065   &   0.9195 & 0.9192 & 0.8248 \\
 7 & 0.9066 & 0.7857 & 0.6311   &   0.9471 & 0.9468 & 0.8524 \\
 8 & 0.9301 & 0.8214 & 0.6510   &   0.9652 & 0.9649 & 0.8705 \\
 9 & 0.9478 & 0.8481 & 0.6678   &   0.9771 & 0.9769 & 0.8825 \\
 10 & 0.9608 & 0.8687 & 0.6831  &   0.9850 & 0.9847 & 0.8903 \\
\end{tabular*}
\caption{Cross: convergence of probabilities under protocol proposed in Section \ref{protocolrepeatedII} compared with simple repetition, as discussed in Section \ref{protocolrepeatedI}.}\label{resultstablecomparisoncross}
\end{table*}
\begin{table*}
\begin{tabular*}{10cm}{@{\extracolsep{\fill}}c|c c c|c c c}
 & \multicolumn{3}{c|}{Protocol II} & \multicolumn{3}{c}{Protocol I} \\
 Measurement & $N=4$ & $N=8$ & $N=12$ & $N=4$ & $N=8$ & $N=12$ \\
 \hline
        1  &     0.4998 & 0.4658 & 0.1586 &       0.4998 &   0.4658 &  0.1586 \\
       2  &     0.7333 & 0.5566 & 0.2485 &    0.7498 &   0.7146 &   0.2920 \\
       3  &     0.8578 & 0.6085 & 0.3234 &    0.8748 &   0.8475 &   0.4042 \\
       4 &      0.9242 & 0.6522 & 0.3959 &    0.9374 &   0.9186 &   0.4987 \\
       5  &     0.9596 & 0.7056 & 0.4623 &    0.9687 &   0.9565 &   0.5782 \\
       6   &    0.9785 & 0.7350 & 0.5179 &    0.9843 &   0.9768 &   0.6451 \\
       7    &   0.9885 & 0.7740 & 0.5676 &    0.9922 &   0.9876 &   0.7013 \\
       8     &  0.9939 & 0.8061 & 0.6128 &    0.9961 &   0.9934 &   0.7487 \\
       9      & 0.9967 & 0.8424 & 0.6509 &    0.9981 &   0.9965 &   0.7885 \\
      10      & 0.9983 & 0.8646 & 0.6863 &       0.9990 & 0.9981  &   0.8221 \\
\end{tabular*}
\caption{Loop: convergence of probabilities under protocol proposed in Section \ref{protocolrepeatedII} compared with simple repetition, as discussed in Section \ref{protocolrepeatedI}.}\label{resultstablecomparisonloop}
\end{table*}
\end{small}

We have calculated the quantity $P_n(t_1,\ldots,t_n)$ for various values of $n$ and $N$, and found that this quantity does indeed converge to unity, but much more slowly than the simple reptition proposed in Section \ref{protocolrepeatedI}.  For small systems though, the rate of convergence using the two protocols is comparable (see Tables \ref{resultstablecomparisoncross}, \ref{resultstablecomparisonloop}); however, the protocol based on the conditional resetting of the system has the obvious advantage that the system does not need to be reset at each stage.

We noted above that there was an initial peak in the success probability, after which the probability fell substantially.  Such a peak becomes much diminished on subsequent measurements, causing the convergence of the success probability to slow as the excitation disperses over the system.

\section{Summary}
We have proposed a system that performs {\em both} the creation
and distribution of entanglement.  These tasks are fundamental to
any physical realization of a quantum computer or quantum
``circuit'', where the ability to create entanglement {\it in
situ} without needing to interfacing different physical systems
would be ideal. Our protocol, for example, could be used to
establish a shared Bell state between two optical lattice quantum
computers or two quantum dot quantum computers without interfacing
atomic systems or quantum dot systems with photons. It is also
directly motivated by schemes of entanglement transfer and
entanglement generation and transfer with minimal control cited in
the introduction. As opposed to the previous protocols of the
latter class, here we conditionally establish a perfect Bell state
between Alice and Bob.

The system conclusively creates a maximally-entangled Bell state
with a certain probability, which varies with the size of the
lattice.  This probability may be improved by repeating the
measurement, or using the more complicated protocol for small
lattices. Advantages of the scheme include the ability to continue
to take measurements without destroying the information, and the
fact that Alice and Bob test for a global state using local
measurements and classical communication.  The only stage that
requires a global action is the resetting of the lattice, though
we have noted that in principle this may also be performed through
local actions.

The probability of success can be slightly lower than hoped, for
larger lattices, but we have shown that it is possible for this to
tend to unity upon repetition, and it may be the case in future
work that the inclusion of the states $\ket{\psi_{2,3}}$ causes
the system to converge without needing to reset. With the
existing protocols we have found that qutrits separated by a
distance of $33$ lattice sites (for a cross of $N=35$) can share a
Bell state with $90$ percent probability of success in just 16 measurements. This might be
a reasonable separation of two distinct quantum processors which
need to be hooked up for greater processing power. 

Our study also provides further insight into the application of SU(3)-invariant Hamiltonians in a quantum information context, which has produced some interesting developments in recent years \cite{article2002bruss-macchiavello, article2002cerf-bourennane-karlsson-gisin, article2003durt-cerf-gisin-zukowski,  article2001spekkens-rudolph}.  Further results are expected in this direction, especially in view of the discovery that qutrit implementations optimize the Hilbert space dimensionality \cite{article2004greentree-schirmer-green-hollenberg-hamilton-clark}.  These developments could also lead to novel perspectives concerning the coherent manipulation of quantum information in many-body systems.

\section{Acknowledgments}
C.H. and S.B. acknowledge financial support from the UK Engineering and 
Physical Sciences Research Council through Grants EP/P500559/1 and 
GR/S62796/01, respectively.  This research is part of QIP IRC  
www.qipirc.org (GR/S82176/01), through which A.S. is supported.  We 
thank Kurt Jacobs and Vladimir Korepin
for very useful discussions, and Daniel Burgarth for pointing out that 
we can reset the system through local operations.

\appendix
\section{Symmetry of Bell state on cross}
Since our aim is to create a maximally entangled bipartite state, it could be asked why we are considering the state $\ket{\psi^+_{AB}}$ alone, and not the more general state
\begin{align}
\ket{\psi^\phi_{AB}}= \frac{1}{\sqrt{1+|r|^2}} \left[ \ket{+1}_A\ket{-1}_B + r\me^{\mi\phi}\ket{-1}_A\ket{+1}_B\right]. \label{bellgeneral}
\end{align}
However, by considering the symmetry of the graph, it is clear that the probability of this state is always zero unless $\phi = 0$.  

In fact, the dynamics are manifestly symmetric under the exchange of Alice and Bob's site, and between sites 1 and 2.
Suppose we have the evolution
\begin{eqnarray}
\ket{+1}_1 \ket{-1}_2 \stackrel{U}{\to} \ket{+1}_A\ket{-1}_B + r\me^{\mi\phi}\ket{-1}_A\ket{+1}_B.  \label{symm1}
\end{eqnarray}
Now let us permute the indices of sites 1 and 2 (or equivalently rotate the graph around the Alice\ndash Bob axis) \footnote{Since the sites are distinguishable, we do not need to worry about the accumulation of a phase here.}, and then invoke the symmetry of the group by permuting the states $\pm 1$.  We then have:
\begin{eqnarray}
\ket{+1}_1 \ket{-1}_2 \stackrel{U}{\to} \ket{-1}_A\ket{+1}_B + r\me^{\mi\phi}\ket{+1}_A\ket{-1}_B.  \label{symm2}
\end{eqnarray}
Since the labelling must not affect the situation physically, we can see by comparing equations (\ref{symm1}) and (\ref{symm2}) that we must have $r\me^{\mi\phi}=1$.

\section{Commutation relations}
We asserted above that the SU($N$) algebra (\ref{algebra}) is satisfied by both fermionic and bosonic creation and annihilation operators.  Here we shall verify this general result.

By definition bosons satisfy the commutation relations $[b^i,b^j]=0$, $[b^\dag_i,b^\dag_j]=0$, $[b^i,b^\dag_j]=\delta_{ij}$ and for fermions we have the anti-commutation relations $\{c^i,c^j\}=0$, $\{c^\dag_i,c^\dag_j\}=0$, $\{c^i,c^\dag_j\} = \delta_{ij}$.

Let us define operators $S^\beta_{\alpha,i}=b^\dag_{\beta,i}b^{\alpha,i}$ and $S^\beta_{\alpha,i}=c^\dag_{\beta,i}c^{\alpha,i}$ in the bosonic and fermionic case, respectively (where $i$ refers to the species, and $\alpha,\beta$ to the state).  Making use of the standard commutation and anti-commutation relations, we find \footnote{We have used $[AB,C]=[A,C]B+A[B,C]$ and $[A,BC]=[A,B]C+B[A,C]$, and for the fermionic case we have also used $[A,BC]=\{B,A\}C-B\{C,A\}$.} the following: 
\begin{widetext}
{\it Bosons:}
\begin{align}
\left[S^\beta_{\alpha,i},S^\rho_{\sigma,j}\right] &= \delta_{ij} \left[b^\dag_{\beta,i} b^{\alpha,i},b^\dag_{\rho,i}b^{\sigma,i}\right]\\
&= \delta_{ij}\left\{ \left[b^\dag_{\beta,i},b^\dag_{\rho,i}\right]b^{\sigma,i}b^{\alpha,i} + b^\dag_{\rho,i}\left[b^\dag_{\beta,i},b^{\sigma,i}\right]b^{\alpha,i} + b^\dag_{\beta,i}\left[b^{\alpha,i},b^\dag_{\rho,i}\right]b^{\sigma,i} +   b^\dag_{\beta,i}b^\dag_{\rho,i}\left[b^{\alpha,i},b^{\sigma,i}\right] \right\}\\
&= \delta_{ij}\left\{ -b^{\dag}_{\rho,i}\delta_{\beta\sigma}b^{\alpha,i} + b^\dag_{\beta,i}\delta_{\alpha\rho}b^{\sigma,i}\right\}\\
&= \delta_{ij}\left\{ \delta_{\alpha\rho}S^\beta_{\sigma,i}-\delta_{\beta\sigma}S^\rho_{\alpha,i}\right\}\qed
\end{align}
{\it Fermions:}
\begin{align}
\left[S^\beta_{\alpha,i},S^\rho_{\sigma,j}\right] &= \delta_{ij}\left[c^\dag_{\beta,i}c^{\alpha,i},c^\dag_{\rho,i}c^{\sigma,i}\right]\\
&= \delta_{ij}\left\{ \left[c^\dag_{\beta,i},c^\dag_{\rho,i}c^{\sigma,i}\right]c^{\alpha,i} + c^\dag_{\beta,i}\left[c^{\alpha,i},c^\dag_{\rho,i}c^{\sigma,i}\right]\right\}\\
&= \delta_{ij} \left\{ \left\{c^\dag_{\rho,i},c^\dag_{\beta,i}\right\}c^{\sigma,i}c^{\alpha,i} - c^\dag_{\rho,i}\left\{c^{\sigma,i},c^\dag_{\beta,i}\right\}c^{\alpha,i} + c^\dag_{\beta,i}\left\{c^\dag_{\rho,i},c^{\alpha,i}\right\}c^{\sigma,i} - c^\dag_{\beta,i}c^\dag_{\rho,i}\left\{c^{\sigma,i},c^{\alpha,i}\right\} \right\}\\
&= \delta_{ij} \left\{ -c^\dag_{\rho,i}\delta_{\sigma\beta}c^{\alpha,i} + c^\dag_{\beta,i}\delta^{\rho\alpha}c^{\sigma,i} \right\}\\
&= \delta_{ij} \left\{ \delta_{\rho\alpha}S^\beta_{\sigma,i} - \delta_{\sigma\beta}S^\rho_{\alpha,i} \right\}\qed
\end{align}\end{widetext}


\end{document}